\documentclass[runningheads]{svmult}

\usepackage{makeidx}   
\usepackage{graphicx}  
\usepackage{aas_macros}
\usepackage{amsmath}
\usepackage{amssymb}

\begin{document}
\title*{Supernova Explosions from Accretion Disk Winds}
\toctitle{Supernova Explosions from Accretion Disk Winds}
\titlerunning{Wind-Driven Supernovae}
%
\author{A. I. MacFadyen}

\authorrunning{A. I. MacFadyen}
%
%
\institute{Theoretical Astrophysics, California Institute of Technology, Pasadena, CA 91125 USA}

\maketitle              

\begin{abstract}

Winds blown from accretion disks formed inside massive rotating stars
may result in stellar explosions observable as Type Ibc and Type II
supernovae.  A key feature of the winds is their ability to produce
the radioactive $^{56}$Ni necessary to power a supernova light curve.
The wind strength depends on accretion disk cooling by neutrino
emission and photo-disintegration of bound nuclei.  These cooling
processes depend on the angular momentum of the stellar progenitor via
the virial temperature at the Kepler radius where the disk forms.  The
production of an observable supernova counterpart to a Gamma-Ray Burst
(GRB) may therefore depend on the angular momentum of the stellar
progenitor.  Stars with low angular momentum may produce a GRB without
making an observable supernova.  Stars with large angular momentum may
make extremely bright and energetic supernovae like SN 1998bw.  Stars
with an intermediate range of angular momentum may simultaneously produce a
supernova and a GRB.

\end{abstract}

\section{Introduction}

Observational evidence continues to establish the association of long
gamma-ray bursts with active star forming regions of galaxies e.g.,
\cite{AF02}.  In addition, there are indications from the optical
light curves of several (long) GRB afterglows that supernova components
may be directly observed as emission from the decelerating
relativistic ejecta fades \cite{Blo99}\cite{Rei99}\cite{Blo02a}.  These
observational continue to support the collapsar models for GRBs in
which the core of a massive rotating star collapses to a black hole
and rapidly accretes \cite{SW93}\cite{MW99}.  It is therefor of
interest to try to understand under what conditions a star which makes
a GRB will also make an observable supernova.

Collapsars \cite{MW99}\cite{MWH01} form dense accretion disks ($\rho
\gtrsim 10^9$ gm cm$^{-3}$) which are extremely optically thick to
photons ($\tau_{\gamma} \sim 10^{19}$).  As the stellar gas spirals
through the disk, photons are trapped and accrete with the gas.  This
is in distinction from ``thin''accretion disks in which photons are
assumed to escape to infinity carrying away the locally dissipated
energy.  Since photons are trapped, viscous dissipation of orbital
energy increases the disk entropy, pressure gradients are important
for the force balance and the disk is ``thick.'' Such non-radiating
accretion flows are capable of ejecting gas away from the black hole
\cite{HBS01}.  Accretion in these disks is inefficient with
significant fractions of the gas supplied at large radii being ejected
from the system.

An important feature of collapsar disks, is the realization at
sufficiently high accretion rates ($\dot{M} \sim 0.1 M_{\odot}
s^{-1}$) of temperatures ($T \sim 10^{10}$ K) and densities ($\rho
\sim 10^9$ gm cm$^{-3}$) at which the loss of thermal energy to
neutrino emission and photodisintegration of heavy nuclei allows for
accretion with a range of efficiency.

\section{WINDS}

MacFadyen \& Woosley \cite{MW99} showed that collapsar disks eject
comparable amounts of material in a wind as is accreted by the central
black hole.  The fraction of accreted gas depends on the efficiency of
neutrino cooling at removing entropy from the accreting gas.  The
remainder is ejected from the black hole as an outflowing wind.
Recent semi-analytic work \cite{NPK01} has mapped the parameter space
of inefficient neutrino-cooled accretion in agreement with detailed
calculations of \cite{MW99} for limited parameters.

Of particular interest in the case of collapsars is the chemical
composition of the wind.  Collapsar disks are hot enough to completely
photodisintegrate heavy nuclei to free nucleons (neutrons and
protons).  Recent simulations \cite{MW02} and \cite{MW99} show
expulsion of free nucleon gas in the wind.  This is important for two
reasons: 1) energetics: free nucleons combining to iron group nuclei
(e.g. Nickel-56) release 8 MeV/nucleon or $1.5 \times 10^{52}$ erg per
solar mass of recombined material.  2) observability: this ejected
material provides a long term energy supply to the explosion
(through radioactive decay of $^{56}$Ni) enabling the gas to shine on
time scales of months.

\section{Light Curve}

Models of the light curve of the energetic and peculiar Type Ibc
supernova SN1998bw require large quantities of $^{56}$Ni (M($^{56}$Ni)
$\sim 0.5 M_{\odot}$) \cite{Iwamoto98}\cite{WES99}.  Conventional
models require large explosion energies to produce sufficient nickel
and fit the light curve.  In addition abnormally high expansion
velocities were inferred from the unusual spectrum indicating a large
explosion energy ($\sim 10^{52}$ erg).  Several groups have also interpreted
deviations from power law decay of GRB optical transients as
supernovae light curve components and have matched them with
appropriately shifted 1998bw light curves \cite{Bloom99}.

Since supernovae are invoked to interpret these observations it is
important to note that a stellar explosion (e.g., jets piercing a
star) is not necessarily a supernova.  Supernovae, as an observable
phenomenon, require a persistent source of energy input to power a
light curve for long times (weeks to months).  It is necessary to make
$^{56}$Ni in the explosion so that radioactive decay (to Cobalt to
Iron) injects energy into the gas so that it can shine.  Lacking a
persistent source of energy input, a stellar explosion would be
unobservable via electromagnetic radiation.  Explosion energy released
in the optically thick star would simply be converted to expansion
kinetic energy with little or no light emitted.

In conventional core collapse supernovae some nickel is thought to be
produced via explosive nuclear burning behind the explosion shock.
However, current models for these ``delayed'' supernova explosions
have trouble producing the $10^{51}$ erg for a normal supernova (in
fact, some current models fail to get any explosion at all!) and are
unlikely to be capable of producing the higher energies required for 1998bw.

\section{Angular Momentum}

As we have seen, neutrino cooling and photodisintegration of heavy
nuclei are crucial for allowing gas to accrete efficiently.  The
neutrino cooling depends sensitively on temperature (e.g.,
$Q_{\nu}^{-} \propto T^6$ for neutrino losses due to pair capture on free
nucleons) and therefore on the radius where the disk forms.  This
radius is, in turn, dependent on the angular momentum of the accreting
gas with the disk first forming at the Kepler radius $R_{kep} \equiv
j^2/GM = 2.5 \times 10^7 j_{17}^2 M_3^{-1}$ cm, where $j_{17}$ is the
specific angular momentum of the accreting gas in units of $10^{17}$
cm$^2$ s$^{-1}$ and $M_3$ is the mass of the central black hole in
units of three solar masses.  The virial temperature for gas falling
to its Kepler radius $T_{vir} = GMm_p/3k_BR_{kep} = 3.3 \times 10^{10}
M_3^2 j_{17}^{-2}$ K, where $m_p$ is the proton mass and $k_B$ is the
Boltzmann constant.  In terms of gravitational radii, $R_G
\equiv GM/c^2$, this temperature is $T_{vir} = m_pc^2/3k_Br = 1.8
\times 10^{12} r^{-1}$ K (assuming a Newtonian potential), where $r
\equiv R/R_G$.  We see that gas with $j_{17} \approx 1$ is heated to
above $10^{10}$ K so that it is fully photodisintegrated to free
neutrons and protons from it's original composition of Silicon, Oxygen
and Helium.  This means that capture of electron-positron pairs onto
the free neutrons and protons serves as an efficient neutrino emission
process which cools the gas and helps it to accrete efficiently.  Gas
with $j_{17} \gtrsim 2.6$, however, heats to less then $5 \times 10^9$
K.  At these lower temperatures the heavy nuclei fail to
photodisintegrate and pair capture neutrino cooling is suppressed.
This gas is therefor poorly cooled and subject to being driven from
the disk.  

It is worth noting that gas with $1 \lesssim j_{17} \lesssim 2.6$ is
partially photodisintegrated.  Photodisintegration acts as a loss for
thermal energy for the gas and thus is effectively a cooling process,
robbing about $10^{19}$ erg of thermal energy from every gram of
photodisintegrated nuclei.  This process helps the gas to accrete and
provides free nucleons which enhance the neutrino cooling.

The above discussion assumed $M_3 =1$ though the scaling with $M_3$ is
apparent.

\begin{figure}
  \includegraphics[height=.8\textheight]{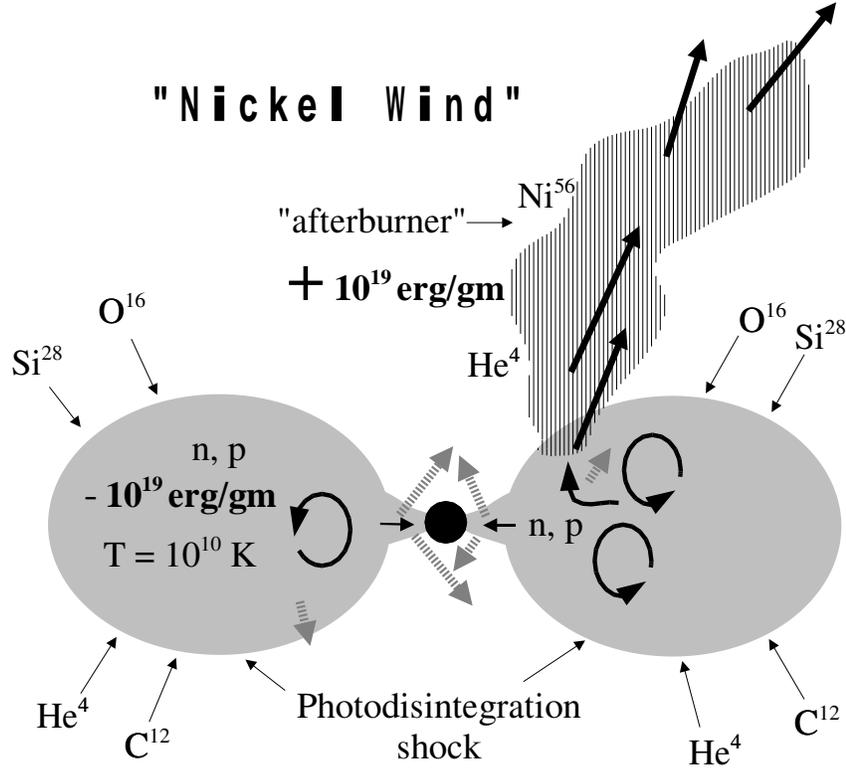} \caption{ Cross section
  of a collapsar accretion disk feeding a stellar mass black hole
  (center).  The disk is embedded in a collapsing star that is falling
  onto the disk at rates above 0.1M$_{\odot}$ s$^{-1}$.  A wind (the
  striped region in the upper right) is blown from the collapsar disk
  at speeds of up to $\sim 40,000$ km s$^{-1}$. The wind is composed
  of free neutrons and protons which can recombine to iron group
  elements injecting $1.5 \times 10^{51}$ erg per $0.1$ M$_{\odot}$ of
  reassembled nucleons.  $^{56}$Ni in the wind can power a long term
  ``supernova'' light curve via radioactive decay of nickel and
  cobalt.  The black solid arrows indicate the velocity of the gas
  flow while the thick dashed lines represent neutrino emission.  The
  wind is shown only in the upper right quadrant for clarity but is in
  reality present in all four quadrants.}

\end{figure}

\section{The Afterburner}

Interestingly, energy lost to photodisintegration becomes available
again if the gas is ejected from the disk and begins to reassemble.

Effectively, gravitational energy is temporarily stored in the freeing
of nucleons from the heavy nuclei.  These nucleons are volatile in the
sense that the have a huge nuclear energy source if they manage to
escape the energetic photons trapped in the (optically thick)
accretion disk.  Accretion physics may provide the nucleons with
opportunity to escape the disk's nuclei-disintegrating photon bath.
Once free they can quickly recover nuclear binding energy by
reassembling into iron group elements.  This process can be explosive
since the nuclei may recombine in seconds compared to millions of years
it took them to assemble (burn) the first time around during the
slower pre-explosion nuclear burning stages.  In fact the reassembly,
plus the kinetic luminosity of the disk wind, may power extremely
energetic explosions.  SN1998bw may be an example.

\section{GRB with Supernova}

The collapsar model relies on rapid accretion into the central black
hole to power relativistic jets which pierce the star and make a GRB
and afterglow via internal and external shocks.  It is notable
that for an interesting range of angular momentum the accretion of the
star simultaneously feeds the black hole rapidly and powers a wind
\cite{MW99}.

There are several interesting regimes determined by the angular
momentum present in the collapsing star:

The following values of angular momentum correspond to important
transition radii in the accretion flow:

$j_{isco}$ angular momentum of the innermost stable circular orbit.
This is the minimum angular momentum needed to form a disk around a
black hole.

$j_{\nu}$ angular momentum of gas that falls deep enough in the
gravitational potential to photodisintegrate the heavy nuclei to free
nucleons activating pair capture neutrino emission as an efficient
coolant.

$j_{\gamma}$ angular momentum of gas that falls deep enough to cool
partially.  Some gas accretes and some is expelled in a wind.  The
relative amount depends on the exact value of $j$

1. $j_{isco} < j < j_{\nu}$ - efficient neutrino cooling allows rapid
accretion into black hole with plenty of power potentially going into jets
with little or no outflows expected.  This kind of star would not be
expected to produce a bright supernova since little or no $^{56}$Ni is
expected to be present in the exploding star.  A possible caveat is
that there is some nickel production via explosion burning in the
lateral jet shock but the temperature is low in this region and not
much mass is involved.

2. $j_{\nu}< j < j_{\gamma}$ - some gas accretes and some is ejected
in a wind rates can be comparable depending on j. This can make both a
GRB and a ``supernova''

3. $j > j_{\gamma}$ Gas doesn't cool efficiently so doesn't feed the
hole rapidly.  Not good for making an accretion powered GRB.  Outflows
may results with some recombination nickel possible if some gas is
heated above $5 \times 10^9$ K by a combination of virialization and
viscous dissipation.  Of interest here is explosive burning of
centrifugally supported oxygen.

A less interesting regime is $j < j_{lso}$ for which the gas falls directly to the innermost stable circular orbit without forming an accretion disk. 

Note that electromagnetic extraction of black hole spin energy is a
possible source of jet energy even for ``slowly'' accreting black
holes.  Convective motions may even be favorable for building up large
magnetic fields needed to extract the hole spin energy.

\section {Photodisintegration}

As stellar gas collapses onto the collapsar accretion disk, adiabatic
compression and shocks can raise the temperature sufficiently to
photodisintegrate the gas to alpha particles and free nucleons.  The
destruction of the heavy nuclei is an energy sink for the gas which is
helpful in allowing accretion to occur.  Photodisintegration of heavy
nuclei (e.g., Silicon, Oxygen in the collapsing core) costs $\eta c^2$
per unit mass where $\eta \approx 0.007$ (or 8 MeV/nucleon) for
complete disintegration to free nucleons.  Gas falling in a
gravitational potential can dissipate its accretion energy by swapping
nuclear binding energy for gravitational binding energy.  A measure of
where this occurs is where the gravitational binding energy equals the
nuclear binding energy $\delta \phi = GM/r = \eta c^2$.  If gas near
the equator falls to it's Keplerian radius the other half goes into
the kinetic energy of Keplerian rotation so $\delta\phi = 1/2 GM/r$.

Measuring radius in gravitational radii $r \equiv \hat{r} r_g$ where
$r_g \equiv GM/c^2$ we can define a photodisintegration radius
$\hat{r}_{pd} \equiv 1/2\eta$.  We thus expect photodisintegration to
``cool'' accreting gas when falls to a radius of $\sim 70\, r_g$.

\section{Viscosity}

The above scenario assumes significant viscosity in the disk gas
corresponding to a Shakura-Sunyaev alpha viscosity parameter $\gtrsim
0.1$.  The temperature and density of the disk wind and hence the
nucleosynthesis depend on the disk viscosity.  Observations of the
supernova powered by a collapsar wind may therefor help to constrain
the viscosity of the collapsar accretion disk.  
             
Recent calculations of the neutron abundance in one-dimensional
accretion disk models relevant to GRBs \cite{PWF99} indicate that the
inner parts of collapsar accretion disks may be too neutron rich to
produce significant quantities of Nickel-56 \cite{PWH02}\cite{AB02}.
Outflows from these inner regions may instead be of interest for rare
nucleosynthesis like the r-process.  However, much of the mass loss
from GRB accretion disks (collapsars and otherwise) may come from the
outer regions of the disks where electron capture has not
significantly neutronized the material composing the outflow.

Collapsar disks may have more than one active wind blowing region:
1. the outer disk where low densities imply non-degenerate electrons
and little neutron excess ($Y_e\sim0.5$).  Radioactive Nickel-56 may
result from nucleosynthesis taking place in as this wind expands.  2.
The innermost disk where the disk becomes optically thick to neutrino
emission and is again poorly cooled.  This innermost disk wind is
significantly neutronized and will not produce Nickel-56.  However, it
is an interesting site for the r-process because of the large neutron
fraction and large entropies attained from viscous dissipation.

In between these two disk regions, the disk is (partially)
neutrino-cooled and most (but not necessarily all) of the gas can
accrete.  This region of the disks transports the fraction of gas
received from the outer, i.e. the gas not ejected in the wind, plus gas
falling onto the neutrino-cooled region of the disk.

\end{document}